\documentstyle[twoside,fleqn,espcrc2]{article}

\newcommand{\AmS}{{\protect\the\textfont2
   A\kern-.1667em\lower.5ex\hbox{M}\kern-.125emS}}

\title{Soft Supersymmetry Breaking and the Supersymmetric Standard
Model\thanks{Invited talk presented at the ``Thirty Years of Supersymmetry"
Symposium,
         University of Minnesota, October 13-15, 2000.}}

\author{Savas Dimopoulos\address[SU]{Physics Department, Stanford
University, Stanford, CA 94305-4060, USA.}
         }

\begin{document}

\begin{abstract}
We recall how the idea of Softly Broken Supersymmetry led
to the construction of the Supersymmetric Standard Model in
1981. Its first prediction, the supersymmetric unification
of gauge couplings, was conclusively verified by the LEP
and SLC experiments 10 years later. Its other predictions
include: the existence of superparticles at the electroweak
scale; a stable lightest superparticle (LSP) with a mass of
$\sim 100$ GeV, anticipated to be a neutral electroweak
gaugino; the universality of scalar and gaugino masses at
the unification scale. The original motivation for the
model, solving the hierarchy problem, indicates that the
superparticles should be discovered at the LHC or the
TeVatron.

\vspace{1pc}
\end{abstract}

\maketitle
\section{Introduction}

It is a pleasure to recall the ideas that led to the
Supersymmetric Standard Model. Supersymmetry is a marvelous
theoretical idea whose mathematical foundations originate
in the early the '70s\cite{SUSY,sugra}. In spite of this,
it took a decade before a potentially realistic theory, one
that is not obviously wrong, was proposed in 1981. The
basic ingredient, missing until that time, was the concept
of Softly Broken Supersymmetry. This is analogous to the
history of the standard model whose mathematical
foundations were laid down by Yang and Mills in the '50s,
but whose development had to wait until the '60s; the
missing idea in that case was that of spontaneous symmetry
breaking.

In this talk I will mostly concentrate on my papers with
Howard Georgi \cite{dg} from the spring of 1981 in which
the idea of Soft Supersymmetry Breaking was proposed and
used to construct what is now called the Supersymmetric
Standard Model (SSM)as well as its unification into $SU(5)$
\cite{GUTs}. The catalyst for our work was the hierarchy
problem\cite{hier,hierSUSY}. At present, the main reason
why the unified Supersymmetric Standard Model enjoys its
status as the leading contender for physics beyond the
standard model is a quantitative {\bf prediction}, dating
from this paper, that has been verified by high precision
data: that is a correlation between $\alpha_s(M_Z)$ and
$\sin^2(\theta_W)$ which has been confirmed by experiment
at the 1\% level \cite{lp} and shows that in the presence
of superparticles at around a $TeV$ the gauge forces of
nature unify at a scale of $\sim 2\times 10^{16} GeV $. In
fact this is the only significant {\it quantitative}
success of any extension of the standard model and
--together with neutrino masses-- is the strongest experimental
indication for new physics. The success of this prediction
depends crucially on having both Unification {\bf and} low
energy Supersymmetry in the same theory; either Unification
or Supersymmetry alone are insufficient. So, although we
have not seen any real superparticles yet, we have evidence
for Supersymmetric Unification via the effects of virtual
superparticles running around loops at energies between the
weak and the unification scale!

  We present the developments in
chronological order, beginning with an overview of the
status of model building before 1981 (section 2). In
section 3 we present the early work with S. Raby and F.
Wilczek on the supersymmetric unification scale and the
absence of proton decay in supersymmetric theories. Section
4 focuses on the papers with H. Georgi where we introduced
the soft terms and the basic ingredients of the
supersymmetric standard model, including the supersymmetric
unification prediction. Section 5 deals with some of the
important theoretical developments that followed. Section 6
discusses the significance of the unification prediction
and its implications for model building and string theory.
We end with an evaluation of the present status of the
Supersymmetric Standard Model in section 7.

\section{Before  1981.}

{\bf Hierarchy Problem:} A crucial turning point in our
field occurred in the Spring of 1978. The SLAC experiment
on parity violation in neutral currents convinced many
theorists that the Standard Model of Glashow, Weinberg and
Salam was correct and that it was a good time to start
focusing on the next layer of questions: to explain some of
the puzzling features of the Standard Model. The first
question that theorists turned to was the ``hierarchy
problem'' \cite{hier}: attempting to understand why the
Higgs mass is so much smaller than the Planck mass or the
Unification Scale. The Higgs does not carry any symmetry
that ensures its lightness; indeed, in the absence of
miraculous cancellations, the Higgs mass would be driven to
the Planck or unification scale; it would not be available
at low energies to do its intended job of giving mass to
the weak gauge bosons and fermions.

  Susskind and Weinberg \cite{ws} proposed the very appealing idea of
Technicolor, as an alternative to the Higgs, for giving mass to
the weak gauge bosons. In early '79 Technicolor was enlarged into
``Extended Technicolor'' \cite{etc} to allow the quarks and
leptons to get their masses. By the summer of 1980 it became clear
that these theories suffered from generic problems of flavor
violations \cite{je} that could perhaps be cured only by
complicating the theory immensely and losing any hope of
calculability. I, perhaps prematurely, felt that this was too high
a price to pay and decided to look at other alternative approaches
to the Hierarchy problem.

That is when we turned to Supersymmetry \cite{SUSY,sugra}.
It was generally realized that Supersymmetry could help the
hierarchy problem \cite{hierSUSY}. The reason is that the
Higgs, a scalar, would form a degenerate pair with a
fermion, called the Higgsino. Since the Higgsino could be
protected by a chiral symmetry from becoming superheavy, so
could its degenerate scalar partner, the Higgs. Of course
Supersymmetry does much more than to just relate the Higgs
to the Higgsino. It assigns a degenerate scalar
``superpartner'' to each and every known quark and lepton,
as well as a fermionic degenerate superpartner to each
gauge boson. Since no such particles had been seen it was
clear that Supersymmetry had to be a broken symmetry.
Nevertheless, Supersymmetry would still help the hierarchy
problem as long as its breaking occurs near the weak scale.
This had the immediate implication that the superpartners
had to be at accessible energies! This line of reasoning
led us to begin our attempt to find a Supersymmetric
version of the Standard Model with Supersymmetry broken at
the weak scale. Together with Stuart Raby and Leonard
Susskind we started learning about Supersymmetry and tried
to find out if such theories had already been constructed.
We quickly discovered that no Supersymmetric versions of
the Standard Model existed at that time.

{\bf Broken Charge and Color:} There were early attempts by
Fayet \cite{fayet} to build models where supersymmetry was
broken spontaneously in the standard model sector. They
were all plagued by a plethora of problems including: the
breaking of electromagnetic gauge invariance, predicting a
photon mass $ \sim 100 ~GeV$; the breaking of color
symmetry at the electroweak scale; massless gluinos, a
consequence of the problematic continuous R-symmetry of
these models. Attempts to cure these problems by enlarging
the gauge group to led to anomalies whose cure again led to
the breaking of the electromagnetic and color gauge
invariance. The root of these problems was that in these
theories Supersymmetry was broken spontaneously at the tree
level. In 1979 a very important paper by Ferrara,
Girardello and Palumbo \cite{fgp} showed that in such
theories, under very general conditions, color and charged
scalars would get negative masses squared, leading to
breaking of electric charge and color. This essentially
stopped efforts to build realistic Supersymmetric theories.
It was hard to take seriously theories in which photons and
gluons weighed $\sim 100$ GeV.

{\bf Supercolor:} We spent early fall of 1980 rediscovering
these problems. It rapidly became clear that {\it the
breaking of supersymmetry had to originate outside the
standard model.} Our first attempt was to break
Supersymmetry dynamically with a new strong force, very
similar to Technicolor, which we called Supercolor. We were
not alone in these efforts. Witten \cite{hierSUSY} as well
as Dine, Fischler and Srednicki \cite{hierSUSY} were
pursuing similar ideas for precisely the same reasons. They
wrote two very important papers entitled ``Dynamical
breaking of Supersymmetry ''(Witten) and ``Supersymmetric
Technicolor'' (Dine, Fischler and Srednicki). Their
preprints appeared in April of '81 at the same time as our
``Supercolor'' paper \cite{hierSUSY}.

  An essential objective of these works was to point out that
low energy Supersymmetry helps the hierarchy
problem\footnote{Lots of people, in addition to those in
Reference \cite{hierSUSY}, were aware of this. The
challenge was to implement the idea in a consistent theory
with weak-scale Supersymmetry .}, and to argue that a new
strong force analogous to QCD or Technicolor may induce the
breaking of Supersymmetry and explain the smallness of the
electroweak scale. Dine, Fischler and Srednicki, as well as
Raby and myself, also attempted to build explicit models
incorporating these ideas, but without much success. I do
not have time to discuss these ``Supercolor'' or
``Supersymmetric Technicolor'' theories. They had problems;
one of them was that they were baroque. By January of 1981
we were
  discouraged. Although Stuart Raby and I had begun writing the
Supercolor paper \cite{hierSUSY}, we already did not believe in
it. It seemed too much to believe that Nature would make
simultaneous use of Supersymmetry {\bf and} Technicolor to solve
the hierarchy problem.

\section{``Supersymmetry and the Scale of Unification.''}

In spite of these obstacles, we were mostly optimistic that
the problem of supersymmetry breaking would eventually be
solved. In the meantime we were getting anxious to start
doing physics with the idea of weak-scale supersymmetry. A
result of this was the early paper with Stuart Raby and
Frank Wilczek \cite{drw1} in which we computed the
Unification Mass in the presence of the minimal
Supersymmetric particle content at the weak scale. We found
that, because the superpartners of the gauge bosons slow
down the evolution of the couplings, the unification mass
increased to about $10^{18}$ GeV. This was interesting for
two reasons:
\begin{itemize}
\item This value is close to the Planck mass, perhaps suggesting
eventual unity with gravity. This connection subsequently
got weaker as more accurate calculations \cite{dg,lp,sin2}
reduced the value to $\sim 2\times 10^{16}\ GeV$.
\item There was a distinct experimental difference with ordinary
$SU(5)$: the  proton lifetime was unobservably long.
\end{itemize}
The latter appeared to be an easily disprovable prediction.
In fact by that time three different experimental groups
had reported preliminary proton decay ``candidate events'':
the Kolar gold field, Homestake mine and the Witwatersrand
experiments. We knew that S.Miyake, of the Kolar Gold Field
experiment, and possibly representatives of the other
experiments were going to talk about their events in the
upcoming ``Second Workshop on Grand Unification'' where I
was also going to present our theoretical
results\cite{swogu}. So, I was a bit nervous but did not
hesitate to present them. I am proud of this paper: A
simple and well motivated ingredient, virtual
TeV-superparticles, made a big difference to a quantity
that was being measured at that time, the proton lifetime.
Perhaps this is the first test that supersymmetric unified
theories (SUSY-GUTs) have passed. In this paper, although
we pointed out that the value of $\sin^2(\theta_W)$ would
change due to the Higgs sparticles, we did not present the
new value. After satisfying ourselves that it would not be
grossly modified, we focused on the change in the
unification mass, which at that time was more important for
experiment.

The next big step was to construct a realistic supersymmetric
theory.

\section{``Softly Broken Supersymmetry and SU(5).''}

{\bf Soft Supersymmetry Breaking:}~In the meantime, the
problem of supersymmetry breaking continued to be a major
obstacle to building a realistic supersymmetric extension
of the standard model. After finishing the previous paper
we, in collaboration with Howard Georgi, returned to this
problem. The prevailing view at that time was that a
realistic Supersymmetric model would not be found until the
problem of Supersymmetry Breaking was solved. It was
further believed that the experimental consequences of
Supersymmetric theories would strongly depend on the
details of the mechanism of Supersymmetry breaking. After
all, it was this mechanism that caused the phenomenological
disasters of the early attempts.

The key that took us out of this dead end was the
realization that {\it a search for a detailed mechanism of
supersymmetry breaking might be futile, unless it also
solves the cosmological constant (CC) problem. Any
mechanism that fails to do this appears so massively wrong
that it seems pointless to trust its secondary
implications, such as its spectroscopy.} This, admittedly
idealistic view, led us to seek a far more general approach
to supersymmetry breaking; one which would have a
  better chance of adapting to describe the effect on
the standard model superparticles of the
--still unknown-- ``correct supersymmetry breaking mechanism''
which must solve the CC problem.

This thought naturally focused us on the standard model
sector and led us to the simplest hypothesis: to start with
a supersymmetric version of the standard model and just add
all the terms which break supersymmetry ``softly''. Our
definition of ``softly'' was dictated by our desire to
address the hierarchy problem: it meant that supersymmetry
breaking went away rapidly enough at high energies that it
did not cause any quadratic divergences to the Higgs mass.
The virtue of this simple effective-field-theoretic
approach is that it is general enough to have a chance of
adapting to the correct ultimate mechanism.

It has some immediate physics implications, since it
implicitly postulates that the dynamics that breaks
Supersymmetry is external to the ordinary $SU(3)\times
SU(2)\times U(1)$ sector; specifically, it implies that:
\begin{enumerate}
\item The only particles carrying $SU(3)\times SU(2)\times
U(1)$ quantum numbers are the ordinary ones and their
Superpartners that reside at the weak scale. Extra
particles with exotic $SU(3)\times SU(2)\times U(1)$
quantum numbers are unnecessary. This is essential for the
successful unification prediction.
\item Ordinary particles and
their superpartners do not carry any extra new gauge
interactions at low energies. This too is important for the
unification prediction.
\end{enumerate}
In summary, the successful gauge coupling unification is
evidence in favor of these two implications of the
hypothesis of soft supersymmetry breaking. The hypothesis
postulates that the origin of susy breaking lies outside
the standard model particles and therefore leaves the
standard model degrees of freedom as simple as can be.

{\bf Main Results:} ~The hypothesis of soft susy breaking
immediately led to the two papers \cite{dg} entitled
``Softly Broken Supersymmetry and SU(5)'' and ``
Supersymmetric GUTs'' which first proposed the
supersymmetric standard model. More precisely, these papers
accomplished three objectives:

\begin{enumerate}
\item {\bf Supersymmetric Unification (SUSY-GUTs):}
  Construction of a Unified supersymmetric theory of strong and
electroweak forces. Our gauge group was SU(5). Unification
was essential for the prediction of $\sin^2(\theta_W)$ and
for some of the phenomenology, such as proton decay and
gaugino masses. It was also important for addressing the
hierarchy problem and related issues such as
doublet-triplet splitting.
\item {\bf Supersymmetry Breaking:}
Supersymmetry was broken softly by mass terms for all
scalar superpartners and gauginos. The origin of
supersymmetry breaking lay outside the standard model
degrees of freedom, as explained earlier in this section.
``Softly'' ensured that the Higgs mass had no quadratic
sensitivity on the unification mass.
\item {\bf Supersymmetric} {\bf Standard Model}:
As a bonus, our theory contained the first
phenomenologically viable supersymmetric extension of the
standard $SU(3)\times SU(2)\times U(1)$ model (SSM),
already imbedded inside the unified theory.

\end{enumerate}
We constructed the model in late March and early April of
1981. We were very pleased. We had the first realistic
Supersymmetric theory, incorporating all non-gravitational
phenomena and valid up to the Planck mass. We immediately
started thinking about experimental consequences. We wanted
to make sure that we would not miss anything important.
Time pressure helped us a lot. Both Howard and I were
scheduled to give two consecutive talks in the Second
Workshop on Grand Unification which took place at the
University of Michigan on April 24-26, 1981 \cite{swogu}.
Here are some of our phenomenological results that we
reported in that Workshop \cite{dg}:
\begin{itemize}
\item {\bf  $\sin^2(\theta_W)$ :}
We presented our SUSY-GUT prediction for
$\sin^2(\theta_W)$. The magnitude we got disagreed with the
then central experimental value, but the errors were large.
We argued that there would have to be 2 Higgs doublets for
the value not to be way off.
\item {\bf Proton Decay}: We reported that the Supersymmetric
Unification Mass is so large \cite{drw1} that proton decay is
unobservably small.
\item {\bf Superparticle Spectroscopy: squarks and sleptons. }
 We noted that if all squarks and sleptons have a common
universal mass ($\sim M_W$) at the unification scale, there
would be a ``Super-GIM mechanism'' supressing neutral
flavor violations. The Higgses could have different masses.
\item {\bf Superparticle Spectroscopy: gauginos. }
Because we had a unified theory all gauginos had a common Majorana
mass ($\sim M_W$) at the unification scale.
\item {\bf Family Reflection Symmetry; Stable LSP. }
To avoid rapid proton decay via dimension-four operators we
postulated a discrete symmetry forbidding three-family
couplings. This symmetry was subsequently called family
reflection symmetry\cite{drw2} or matter parity\footnote{It
turned out to be equivalent to a discrete subgroup of the
problematic continuous R-symmetry\cite{fayet}.}. We
concluded:
\\[2mm]
  {\it ``the
lightest of the supersymmetric particles is stable. The
others decay into it plus ordinary particles. One simple
possibility is that it is the U(1) gauge fermion.''}
\end{itemize}

It is gratifying that the above ingredients have survived
the test of time. They form the basis of what is now called
the minimal supersymmetric standard model (MSSM). Perhaps
the most important conclusion of our paper is also the one
that now seems so evident because it has, with time, been
incorporated into our thinking:
\begin{quote}
{\it ``The phenomenology of the model is simple. In
addition to the usual light matter fermions, gauge bosons
and Higgs bosons, we predict heavy matter bosons, gauge
fermions and Higgs fermions as supersymmetric partners. We
can say little about their mass except that they cannot be
very large relative to 1 TeV or the motivation for the
model disappears.''} \cite{dg}
\end{quote}
Of course, our motivation was to address the hierarchy problem;
without it we could not have drawn this conclusion.

{\bf Early Reception:} ~Georgi and I spoke on the last day
of the conference \cite{swogu}. My feeling then was that
our results were for the most part ignored, especially by
the experimentalists who did not care about the hierarchy
problem. Our conclusions were very much against the spirit
of the conference. There were three things against
us:\\[1mm](1)The central value of the weak mixing angle
agreed better with the predictions of ordinary
(non-Supersymmetric) Grand Unified Theories, albeit with
large error bars.\\(2) Preliminary proton decay ``candidate
events'' had been reported by three different experimental
groups, the Kolar gold field, Homestake mine and the
Witwatersrand experiments.\\(3)The host institution was
gearing up to launch the then biggest effort on
proton-decay, namely the IMB experiment.\\[1mm]The
atmosphere in the conference is summarized by Marciano's
April 24, 1981 concluding remarks \cite{swogu}:\\[1mm]
``The basic idea of Grand Unification is very appealing.
The simplest model based on $SU(5)$ has scored an important
success in predicting a value for {\bf $\sin^2(\theta_W)$
which is in excellent agreement with recent experimental
findings} (after radiative corrections are included). It
makes an additional dramatic prediction that the proton
will decay with a lifetime in the range of
$10^{30}$--$10^{32}$ years. If correct, such decays will be
seen by the planned experiments within the coming year
({\bf or may have already been seen}). An incredible
discovery may be awaiting us.''\footnote{The emphasis here
is mine.}\\[1mm]In spite of this, theorists that cared
about the hierarchy problem were pleased with our work.
This included Sheldon Glashow, Leonard Sus\-skind and
Steven Weinberg. In his April 26, 1981 conference summary
talk \cite{swogu} Weinberg mentioned our theory and its
predictions of $\sin^2(\theta_W)$ and $M_{GUT}$ several
times. Weinberg's verdict \cite{swogu}:
\begin{quote}
{\it ``...the model of Dimopoulos and Georgi has many other
attractive features and something like it may turn out to be
right.''}
\end{quote}

This was music to my ears. In May I presented our results
in two more conferences, one in Santa Barbara and the other
at the Royal Society in London. Soon afterwards theoretical
activity in supersymmetric unification began to pick up. In
August of '81 Girardello and Grisaru wrote a very important
paper \cite{gg} systematically discussing explicit soft
breaking of global supersymmetry; they were the first to
discuss cubic soft terms. Starting in July of '81 several
important papers \cite{sin2} repeated our
  calculation of the superunified value of  $M_{GUT}$ and
$\sin^2(\theta_W)$, some improving it to two loops.
  Sakai's paper \cite{sin2} also repeated our analysis
  of SU(5) breaking; it did not discuss the soft
supersymmetry breaking terms and thus did not address the
spectroscopy and phenomenology of superparticles.

The interest in GUTs and SUSY-GUTs dwindled after 1983. The
rise of superstrings, the absence of proton decay and the
lack of precise data on $\sin^2(\theta_W)$ were some of the
reasons. The best evidence that the morale among the
non-stringers was low is that the annual series of
``Workshops on Grand Unification'' was terminated. 1989 was
the year of the ``Last Workshop on Grand Unification''. In
the introduction to that terminal volume Paul Frampton
exclaimed:\\[1mm] {\it `` Alas, none of the principal
predictions of GUTs have been confirmed.''}\\[1mm] This was
written in August 1989, just as LEP was beginning to take
data...

\section{Completing the Picture.}

Since time is so short I have limited myself to those
aspects of superunified theories that are least
model-dependent and experimentally testable or, in the case
of $\sin^2(\theta_W)$ and proton decay, perhaps already
tested. Of course, the theory that we proposed left some
important theoretical questions unanswered. I will briefly
mention some of the problems and related ideas.

{\bf Proton Decay Revisited:} Although Georgi and I worried
a lot about dimension-four baryon violating operators and
we introduced the family reflection symmetry to forbid
them, it did not occur to us to check the operators of
dimension five! Weinberg \cite{wsy} as well as Sakai and
Yanagida \cite{wsy} studied these operators and concluded
that they pose a severe problem for our theory. They
attempted to construct models with an extra $U(1)'$ gauge
group that would forbid the dimension five operators that
mediated proton decay. Raby, Wilczek and I studied these
operators in October of '81 and concluded that the small
Yukawa couplings of the light generation naturally
supressed them \cite{drw2}. The resulting proton decay
rates, although not calculable from low energy physics
parameters, could be experimentally observable. Furthermore
they had a very unique signature that is not expected in
non-supersymmetric theories: protons and neutrons decay
into kaons. We were very excited that we had identified
another ``smoking gun'' for supersymmetry. Ellis,
Nanopoulos and Rudaz independently reached the same
conclusions \cite{drw2}.

{\bf Doublet-triplet splitting:} There is one remaining
technically natural fine tuning in our theory \cite{dg}.
Wilczek and I addressed this problem in June of 1981 and
found two solutions now called the missing partner and the
missing VEV mechanisms \cite{dw}. Attempts to implement
these mechanisms in realistic theories led to complicated
constructions \cite{gr}.

{\bf Hidden sector:} The theoretical question of how
supersymmetry is broken and superparticle masses are
generated in our theory attracted a lot of attention.
Georgi and I had decided that, in the absence of a solution
to the cosmological constant problem, any specific
supersymmetry breaking mechanism was suspect and should not
be relied upon to predict sparticle masses etc. This was a
reason we proposed our more general soft-terms approach.
Nevertheless, it was important to present at least an
existence proof of a mechanism that generated our soft
terms. An important consideration was that squarks and
sleptons belonging to different generations had to have
identical masses to avoid problems with rare processes
\cite{dg}. In the winter/spring of '82 three different
groups \cite{hs}, Dine and Fischler, Raby and I, and
Polchinski and Susskind came up with the idea of a Hidden
Sector, around $10^{11}$ GeV, where supersymmetry breaking
originates and is subsequently communicated to the ordinary
particles via a new gauge interaction at the unification
scale\footnote{For Raby and me the starting point was
trying to build a realistic model utilizing Witten's idea
of ``Inverted Hierarchy'' \cite{w}. }. Soon afterwards a
series of very important papers developed a better idea for
such a mechanism: Supersymmetry breaking could be
communicated from the hidden sector via supergravity
\cite{suGUT}.

{\bf Radiative electroweak breaking:}  Hidden sector mechanisms
for Supersymmetry breaking, under very special assumptions, give
degenerate masses to all scalars: squarks, sleptons as well as
Higgses. This is good for avoiding flavor violations \cite{dg} but
poses the puzzle: what distinguishes the Higgs from the squarks
and the sleptons? Why does the Higgs get a vacuum expectation
value and not the squarks?\footnote{In the original SUSY-GUT this
was not an issue because the Higgs masses were assumed to be
different from the universal squark and slepton masses
\cite{dg}.}. Starting with Iba\~nez and Ross, a series of very
important papers  \cite{rewsb} developed the idea of radiative
electroweak breaking which answers this question dynamically
provided the top quark is sufficiently heavy, above $\sim 60$
GeV.

The title of this section is misleading. The picture is
still far from complete; many fundamental questions remain
unanswered. The theory we have is definitely {\bf not} a
theory of everything. Instead, it is a phenomenological,
disprovable theory that allows us to make contact with
experiment in spite of the questions that it fails to
address.

\section{ How Significant is the Unification Prediction? }

Since the LEP data confirmed the SUSY-GUT prediction this
topic has received a lot of attention and is discussed in
many papers. My analysis will be somewhat outdated, based
on the excellent analysis of Ref. \cite{lp} and the
overview of ref \cite{erice}. The results have not changed
much since then and supersymmetric unification continues to
be successful.
The estimated uncertainties in the theoretical predictions
for SUSY-GUTs and GUTs are due to: $\alpha_s(M_Z)$ and
$\alpha(M_Z)$ error bars, sparticle thresholds, $m_t$ and
$m_{h^0}$, GUT thresholds and Non-renormalizable operators
at the unification scale. For the $\sin^2(\theta_W)$
prediction they all add up to about $\pm 1\%$
\cite{lp}\footnote{$\sin^2(\theta_W)$ is in the $\overline
{\rm MS}$ scheme. }. The experimental error is negligible,
$\pm 0.2\%$. Experiment and theory agree and the
probability that the agreement is an accident is $\sim
2\%$. The largest source of theoretical uncertainty is due
to the $\alpha_s(M_Z)$ error bar; this should shrink in the
future. The other uncertainties are significantly smaller.
The threshold corrections are proportional to $\alpha$s
times logarithms of mass ratios. For example, the total of
the low energy sparticles' contributions is summarized in
the following expression \cite{lp,cpw}:
\begin{eqnarray}
\sin^2\theta(M_Z) &=& 0.2027 + \frac{0.00365}{\alpha_3(M_Z)}
\nonumber \\ &-& \frac{19 \alpha_{em}(M_Z)}{60 \pi}
\ln\left(\frac{T_{SUSY}} {M_Z}\right)
\end{eqnarray}
where\footnote{In eq.(2) if any mass is less than $M_Z$ it should
be replaced by $M_Z$. },
\begin{eqnarray}
T_{SUSY} &=& m_{\widetilde{H}} \left(
\frac{m_{\widetilde{W}}}{m_{\tilde{g}}} \right)^{28/19}
\left(
\frac{m_{\tilde{l}}}{m_{\tilde{q}}} \right)^{3/19} \times
  \nonumber \\ &&  \times \left(
\frac{m_H}{m_{\widetilde{H}}} \right)^{3/19} \left(
\frac{m_{\widetilde{W}}}{m_{\widetilde{H}}} \right)^{4/19} .
\label{eq:SUSYm}
\end{eqnarray}
and $m_{\tilde{q}}$, $m_{\tilde{g}}$, $m_{\tilde{l}}$,
$m_{\widetilde{W}}$, $m_{\widetilde{H}}$ and $m_H$ are the
characteristic masses of the squarks, gluinos, sleptons,
electroweak gauginos, Higgsinos and  the heavy Higgs doublet,
respectively. $T_{SUSY}$ is an effective SUSY threshold.

 From these equations we learn that the supersymmetric
threshold corrections are typically small. The same holds
for the high energy threshold corrections in minimal
SUSY--GUTs \cite{lp}. Therefore the $\sin^2\theta(M_Z)$
prediction is quite insensitive to the details of both the
low and the high mass-scale physics; it takes a large
number of highly split multiplets to change it appreciably.
For example, we know that to bring $\sin^2\theta(M_Z)$ down
by just $\sim 10\%$ --- back to the standard SU(5) value
--- we would need to lift the higgsinos and the second
higgs to $\sim 10^{14} GeV$.

The flip side of these arguments show that to ``fix''
Standard non-supersymmetric GUTs, you also need several
highly split multiplets \cite{fg}. In fact you need many
more than the supersymmetric case, since you do not have
superpartners. In Standard GUTs either $\sin^2(\theta_W)$
or $\alpha_s(M_Z)$ are off by many standard deviations.
Worse yet, the proton decays too fast. Do these problems
mean that all non-supersymmetric GUTs are excluded? Of
course not. By adding many unobserved split particles at
random to change the running of the couplings you can {\it
accommodate} just about {\bf any} values of
$\sin^2(\theta_W)$ and $M_{GUT}$. So, in what sense are
these quantities {\it predicted}\,?\\ I answer this with a
quote from reference \cite{phystoday}:\\[1mm] `` Once we
wander from the straight and narrow path of minimalism,
infinitely many silly ways to go wrong lie open before us.
In the absence of some additional idea, just adding
unobserved particles at random to change the running of the
couplings is almost sure to follow one of these. However
there are a {\bf few ideas} which do motivate definite
extensions of the minimal model, and {\bf are sufficiently
interesting that even their failure would be worth knowing
about.''}\footnote{Emphasis mine}\\ [1mm]

{\bf Peaceful Coexistence with Superstrings:}\\[0.5mm] The
predictions of the heterotic string theory for
$\sin^2(\theta_W)$ (inputing $\alpha_s(M_Z)$) is off by 26
standard deviations \cite{erice}. Similarly, the prediction
of $\alpha_s(M_Z)$ (inputing $\sin^2(\theta_W)$) is off by
11 standard deviations. The reason is that in the heterotic
models the string scale is rigidly connected to the
observed value of the Planck mass and turns out to be a
factor 20 bigger than the unification scale. As a result,
in heterotic string theory, the predicted value of the
proton mass is $20 GeV$. The reaction of the string
community to this disagreement was mixed. Many celebrated
the indirect evidence for low energy supersymmetry as being
``consistent with string theory''. Some adopted the
attitude that a discrepancy by
  a factor of 20 was not too bad, and chose to ignore that it
was off by a large number of standard deviations. Others
adopted the view that the success of the supersymmetric
unification prediction was an accident and drew parallels
between it and the near equality of the apparent size of
the Sun and the Moon on the sky
\footnote{The success of supersymmetric unification
is now taken more seriously and is the most common
criticism of the large dimension framework \cite{add}.}.
Many found comfort in the possibility that very large
threshold stringy corrections could be tuned to ``fix'' the
problem. Of course, such a ``fix'' is no better than
accommodating ordinary non-supersymmetric SU(5) with large
corrections caused by random unobserved multiplets. The
question remained \cite{barb}:\\[1mm] ``why should these
corrections maintain the relations between the couplings
characteristic of the Grand Unified symmetry, if such a
symmetry is not actually realised?''\\[1mm]This was the
climate until a very important paper by Petr Horava and
Edward Witten \cite{hw} took the supersymmetric unification
prediction seriously and proposed to lower the
  string scale  to match the SUSY-GUT scale
of $\sim 10^{17} GeV$. To explain the weakness of gravity
they proposed a new class of 5-dimensional
  theories in which the relation between the
string scale and the 4-dimensional Planck mass is not
direct but involves the size of the 5th dimension. By
choosing its size large enough, $\sim 10^{-28} cm$, one
could account for the unification of gravity with the other
forces at the now reduced string scale. Although it has not
led to a realistic model, the scenario proposed by Horava
and Witten is a good contemporary example of how input from
experiment can help focus theoretical effort in a new
direction.

\section{An Evaluation of the Supersymmetric Standard
Model}

There is no question that the biggest success of the SSM is
the unification of couplings. Since much of this talk has
been devoted to that, we now want to discuss how well the
SSM does with some other important phenomenological issues.
Many of these are widely viewed as successes of the SSM and
I will attempt to present a more balanced view of the pros
and cons. The second virtue of the SSM --and its original
motivation-- is that it addresses the hierarchy problem, at
least in the sense that it protects light scalars from
ultraviolet physics. This is not quite the same as solving
the hierarchy problem, which requires further dynamics for
obtaining the weak mass from the GUT scale, but it is an
  ingredient ensuring the {\it stability} of the
hierarchy. It is a definite plus, extensively discussed,
and I have nothing to add. The remaining issues, often
considered as virtues of the SSM are: proton longevity,
dark matter LSP, neutrino masses, bottom-tau unification
and approximate neutral flavor conservation. To start with,
these are all qualitative and, as a result, less impressive
than unification. We evaluate them in turn:\\ {\bf Proton
Longevity:} This is a virtue of the {\it non-supersymmetric
and non-unified standard model }, where the conservation of
baryon and lepton numbers is an automatic consequence of
gauge invariance. In contrast, in the supersymmetric theory
we were {\it forced} to introduce an additional global
symmetry, the family reflection symmetry, to account for
the stability of the proton\cite{dg}. Such symmetries are
also necessary in other extensions of the standard model,
such as the large dimension framework \cite{add}. In fact,
the most recent Super-Kamiokande limits to the proton
lifetime are so severe that the dimension five operators of
section 5 may be problematic for simple SUSY-GUTs. One has
to either postulate that the color triplet Higgs-fermions
are significantly heavier than the Planck mass or, more
plausibly, that their vertices have a complicated flavor
structure which comes to the rescue and suppresses the
decay of the proton.\\ {\bf Dark Matter LSP:} The existence
of a stable lightest supersymmetric particle (LSP) as a
dark matter candidate is a welcome qualitative feature of
the SSM. Its stability is a consequence of the
family-reflection-symmetry, postulated to account for the
stability of the proton. This chain of reasoning --new
physics at a TeV requires new symmetries to ensure a stable
proton which in turn implies a new stable particle-- is
common. In the large dimension framework\cite{add} there
are several possibilities for stable DM candidates in the
TeV range, such as matter on other walls or in the bulk.
Furthermore, getting the correct abundance does not require
a miracle. Stable particles in the TeV-range naturally have
the right annihilation cross section to result in remnant
abundance near closure density\cite{ct}.\\ {\bf Neutrino
Masses:} The argument here is that the success of the
seesaw mechanism is an indication for $SO(10)$-like physics
at a large scale scale \cite{gmrs}. Perhaps; but the actual
scale associated with right handed neutrinos is
significantly below the SUSY-GUT scale and the connection
is one of rough orders of magnitude. Furthermore, an
essentially identical
--and equally loose-- connection can be made in the large
dimension framework\cite{bulkneutrinos}. There, neutrino
masses could be argued to give evidence for a large bulk!\\
{\bf Bottom-tau unification:} This too is qualitative, and
works about equally well in the non-supersymmetric standard
model\cite{begn}. Furthermore, this relation fails for the
lighter generations, perhaps because they are more
susceptible to Planckean physics\cite{eg}.\\ {\bf
Approximate Neutral Flavor Conservation.} This in fact is,
just like proton decay, often interpreted backwards: We
were {\it forced to postulate} the universality of scalar
masses to account for the absence of neutral flavor
violations\cite{dg}. One might like to argue in favor of
this on grounds of simplicity. This is obviously not
sufficient since there is no symmetry to ensure the
universality of sparticles masses; the flavor symmetry is
broken badly in the fermion sector and this breaking in
general contaminates the scalar sector and creates unwanted
large flavor violations \cite{hkr}, especially in the kaon
system. The issue of how to avoid this is subtle and has
sparked renewed interest in low-energy-gauge mediated
theories \cite{gm}. There the problem of the contamination
of soft terms by fermion masses is avoided because the soft
supersymmetry breaking vanishes in the UV where flavor
originates.

This is part of the challenging ``Flavor Problem'', one of
the most serious for the SSM: that, even after we impose
all the gauge symmetries (as well as the family-reflection
global symmetry), the model has 125 parameters!
\cite{sutter}. Luckily, the vast majority of these
parameters reside in the flavor sector of the theory and do
not contaminate the successful prediction of the
unification of {\it gauge} couplings.

In summary, the {\it gauge sector} of the SSM is
compelling; the flavor sector requires care to ensure
approximate
 flavor conservation and and proton stability. In contrast,
 the non-supersymmetric  unified theories \cite{GUTs} have problems
 in their gauge sector, both with respect to proton
 decay and gauge coupling unification.

 An often unspoken
practical virtue of the SSM is that it is a perturbative
theory with detailed predictions, for any choice of
parameters. Although this is not fundamental, it accounts
for some of the popularity of the model. This is not the
case for either technicolor or the large dimension
framework, which eventually requires a full string theory
model of the world at a $\sim $ TeV.

Of course, the most serious problem of the SSM is the
cosmological constant (CC) problem. It casts a dark shadow
over everything, including the standard model. It is
possible that all our efforts to go beyond the standard
model based on the hierarchy problem are misguided, because
they have nothing to say about the CC problem. On alternate
days I think this is the right view and that looking under
the hierarchy ``lamp post'' is leading us nowhere. The
other days however I think that we can decouple the CC
problem from the rest, perhaps because it involves gravity.
Or, better yet, because Nature has already told us so, with
the tremendous success of QED and the Standard model. Or,
perhaps even by the very success of the supersymmetric
picture of gauge coupling unification...

Because of my involvement with both the SSM and the large
dimension idea, I am often asked ``which do I believe is
correct''. Obviously, I am not more qualified than anybody
else to answer this question. Still, the unification of
coupling is more natural in the SSM and for this reason I
have a preference for the SSM. However, as I tried to
emphasize in this section, what we do not know far exceeds
what we do. The normal desert picture has, for over $\sim
20$ years, failed to shed light on many questions, such as
the flavor and the CC problems. For these reasons alone it
seems worthwhile to consider alternatives that may provide
a new perspective to old problems.

We are fortunate that in a few years experiment will tell
us which road Nature chooses for breaking the electroweak
symmetry. Either way, we will be living in exciting times.
If it is supersymmetry will see the superpartners. If it is
large dimensions we will see all of quantum gravity and
string theory, so we will have an even more complete
picture of the universe. Or perhaps, best of all,
experiment will tell us something even more strange and
exciting that none of us has dreamed.

\section{Acknowledgments}
I would like to thank Howard Georgi for sharing his
recollections of the events that led us to the
Supersymmetric Standard Model and for a careful reading of
the manuscript. I would also like to thank: N.
Arkani-Hamed, R. Barbieri, M.Carena, G.Giudice, N.Polonsky
and C. Wagner for very valuable discussions; K. Olive, M.
Shifman, S. Rudaz, A. Vainshtein and M. Voloshin for
organizing a stimulating symposium and for their
hospitality. My work, since my student years, has been
supported by the National Science Foundation; present grant
number: NSF-PHY-9870115-003.

\end{document}